\documentclass{iau_FM}
\usepackage{graphicx}
\usepackage{comment}
\usepackage{natbib}
\usepackage{hyperref}

\title[LSSDS. Hands on learning] 
{La Serena School for Data Science and the Spanish Virtual Observatory Schools: \\ Initiatives Based on Hands on Experience }

\author[A. Bayo et al.]   
{A. Bayo$^1$,
 V. Mesa$^{2,3,4}$, G. Damke$^5$, M. Cerda$^8$, M. J. Graham$^6$, D. Norman$^7$ , F. F\"orster$^9$, C. Ibarlucea$^5$ \and N. Monsalves$^{2}$}

\affiliation{$^1$European Southern Observatory, Germany\\ email: {\tt abayo@eso.org} \\[\affilskip]
$^2$Instituto de Investigaci\'on Multidisciplinar en Ciencia y Tecnolog\'ia, Universidad de La Serena\\
$^3$Association of Universities for Research in Astronomy (AURA)\\
$^4$Grupo de Astrofísica Extragal\'actica-IANIGLA, CONICET, Universidad Nacional de Cuyo, Gobierno de Mendoza\\
$^5$Cerro Tololo Interamerican Observatory, NSF’s NOIRLab\\
$^6$California Institute of Technology\\
$^7$NSF’s NOIRLab, Tucson, AZ\\
$^8$Instituto de Ciencias Biom\'edicas \& Centro de Inform\'atica M\'edica y Telemedicina. Facultad de Medicina, U. Chile\\
$^9$Data and Artificial Intelligence Initiative, Center for Mathematical Modeling, U. Chile
}

\pubyear{2024}
\jname{Astronomy in Focus, Focus Meeting 7} 
\editors{Diana M.~Worrall, ed.}
\begin{document}

\maketitle

\begin{abstract}
The worlds of Data Science (including big and/or federated data, machine learning, etc) and Astrophysics started merging almost two decades ago. For instance, around 2005, international initiatives such as the Virtual Observatory framework rose to standardize the way we publish and transfer data, enabling new tools such as VOSA (SED Virtual Observatory Analyzer) to come to existence and remain relevant today.
More recently, new facilities like the Vera Rubin Observatory, serve as motivation to develop efficient and extremely fast (very often deep learning based) methodologies in order to fully exploit the informational content of the vast Legacy Survey of Space and Time (LSST) dataset. 
However, fundamental changes in the way we explore and analyze data cannot permeate in the ``astrophysical sociology and idiosyncrasy" without adequate training. In this talk, I will focus on one specific initiative that has been extremely successful and is based on ``learning by doing": the La Serena School for Data Science. I will also briefly touch on a different successful approach: a series of schools organized by the Spanish Virtual Observatory.
The common denominator among the two kinds of schools is to present the students with real scientific problems that benefit from the concepts / methodologies taught. On the other hand, the demographics targeted by both initiatives vary significantly and can represent examples of two ``flavours" to be followed by others.
\keywords{methods: statistical, astronomical databases: miscellaneous, surveys, education.}
\end{abstract}
\firstsection 

\vspace{-0.1cm}
\section{Introduction}

Much of the light received from the universe is not perceptible to our eyes because it has very long or very short wavelengths. To compensate for this bias, there are astronomical instruments that complement our eyes to reveal the invisible. Even in the optical domain, the same idea of ``encoding" information that we cannot ``see", applies to the advent of large surveys with the ``culminating" data tsunami predicted\footnote{\url{https://dmtn-102.lsst.io/DMTN-102.pdf}} for the Large Synoptic Survey Telescope (LSST) to be conducted by the Vera Rubin Observatory (VRO).

A strategy to acquire the skills necessary to succeed in this highly complex interplay between technical and purely abstract landscapes, can be based on an holistic view of the learning process.

\begin{figure}[h]
\begin{center}
\includegraphics[width=0.85\textwidth]{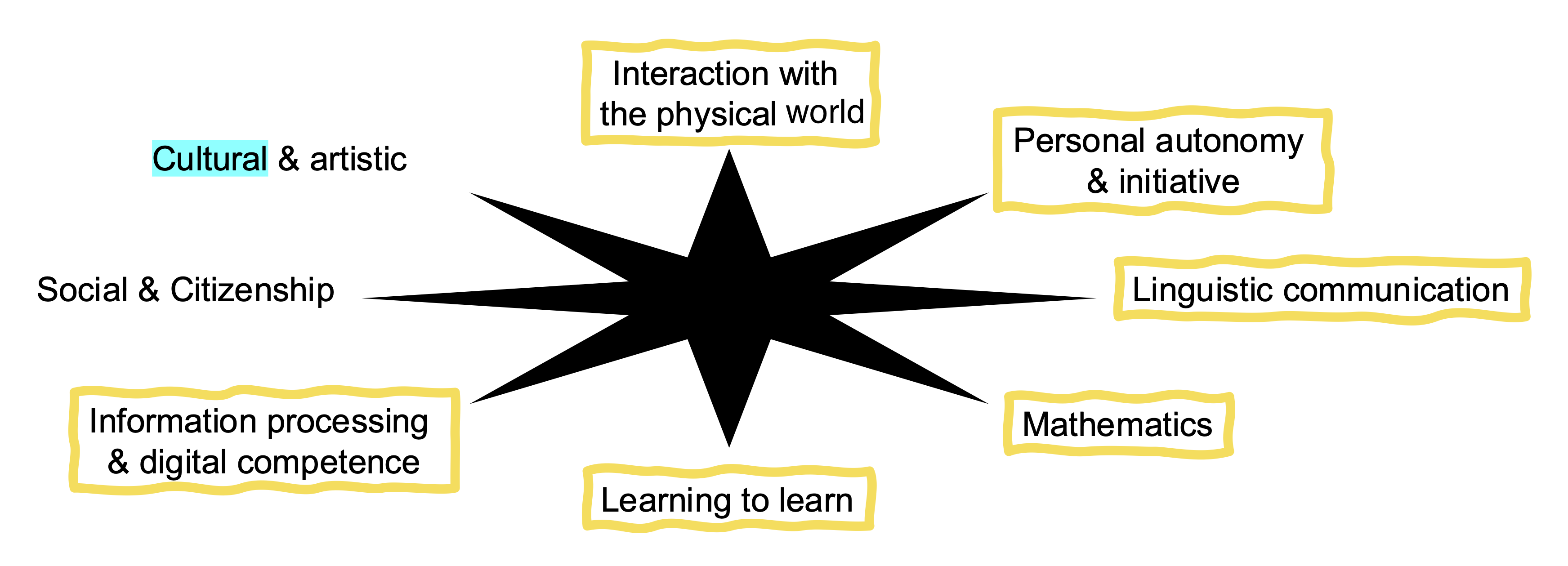} 
\label{competences}
 \caption{Adaptation of the eight basic competences adopted by the Organisation for Economic Co-operation and Development heavily inspired by TMI. Those that consciously (in the last edition) and unconsciously (before) drove our selection criteria for LSSDS students are highlighted. }
   \label{fig1}
\end{center}
\end{figure}

Aligned with this idea, Fig.~\ref{competences} presents an adaptation of the main aspects in Howard Gardner's Theory of Multiple Intelligences (TMI\footnote{\url{https://books.google.de/books?id=bgQ7HzW_3LkC}}). Although emerging from a total different context than that of Data Science, the organizing committee of La Serena School for Data Science (LSSDS) find these competences vital in two aspects: the definition of student selection criteria for our school and our teaching philosophy and curriculum.



\vspace{-.5cm}
\section{The Evolution of the School}
\label{sec:evolution}

The first edition of LSSDS dates back to 2013, with the leadership of the Association of Universities for Research in Astronomy (AURA) and  support from  institutions like the National Science Foundation (NSF) and the Center for Mathematical Modeling (CMM). 

The  motivation behind the school has slightly morphed through the years: at its inception, a clear lack of training in data science related topics was identified in the curriculum taught for physics, astronomy and statistics at both US and Chilean universities (including graduate and undergraduate stages). The teaching style adopted to offer this training has remained constant: a very hands-on, intensive program that promotes team work to tackle realistic research problems.

After three very successful editions (2013-2015), the organizing committee took a one year break to evaluate and identify routes for improvements, as well as to guarantee additional funding. Natural synergies were found with the field of bio-medicine and incorporated at all levels: the faculty, the curriculum and the student recruitment process \citep{Bayo21, Mesa24}. 

Finally, recently, two new challenges had to be faced: the proliferation of online data science related material / courses (of a variety of quality levels, but many lacking not only strong but even light statistical / computer science foundation) and the global pandemic. The former is considered a challenge in the sense that many more students are familiar with machine learning, for instance, but they have not necessarily been provided with critical and solid grounding in. The latter, provoked the cancellation of the 2020 edition of the school and the need to go virtual for two years.

\vspace{-0.5cm}
\section{The selection process}
Despite the changes mentioned in Secs. ~\ref{sec:evolution} and ~\ref{sec:program}, a constant in LSSDS has been the very high level of oversubscription. Between 150 and 300 applications are received for a limited number of $\sim$30 positions. In addition, since we target students with majors or minors in either
mathematics, statistics, physics, computer sciences, astronomy, and bio-related subjects, the diversity of profiles of the applicants is very large. Last, but not least, the school is mostly targeting US and Chilean students, with huge cultural and language differences.

These three factors constantly push the selection committee to carefully consider and try to achieve a complicated balance between fairness and the limited human resources we have to complete the task. 

Regarding representation, we have traditionally included aspects such as gender, field of studies and home university as a ``corrective last step": once a ranking is built (see below), we make the adjustments needed to achieve parity in gender, a flat distribution of fields (which has proven to be very hard as the applicants are heavily biased towards astronomy), to push towards over-representation of ``small" universities, and to prevent that more than one student from a given university is invited. 

The specifics on how the ranking is put together has also morphed through time. However, in rough terms, the ``spirit" has been kept constant: \textit{select the students that can benefit the most from the school and can, hopefully, spread its reach}. This spirit translates into some ``clear" criteria such as: favouring the last two years of undergraduate or the first two years of graduate programs, deprioritizing applications from students that had access to formal training on the topics taught at the school (see Sec.~\ref{sec:program}), but, at the same time, favouring applications where, from the motivation letter, we can see that the students have misconceptions regarding the same topics. However, we also ``fold in" harder criteria to evaluate, such as motivation, future impact in their careers, etc. The evaluation of these aspects with the material we get from the applicants (some data in a web-form, transcripts, a motivation letter and two reference letters) is extremely sensitive to biases, such as language and cultural ones. In recent editions we have tried to address these biases by building a numerical scoring system where skills (those highlighted with yellow boxes in Fig.~\ref{fig1}) are matched to possible expressions, wording, etc proposed by English and Spanish native speakers from different backgrounds. This numerical scoring system was ``anchored" on a small experiment carried out with application material from previous years. This experiment allowed us to confront how, even with metrics designed to be very descriptive and bias-free, the dispersion in the grading of the committee was significant. We understand this as an expression of the diversity in the committee itself, therefore, while vigilant, we kept the same scoring system.

Despite all the efforts previously mentioned to guarantee diversity, for the material to actually reach the students, we request that the students assess their fluency in English (all the material is taught in English) and in at least one programming language. Regarding the latter, most of the school is taught in Python, but the request is considered satisfied with any programming language, as our aim is for the student to have the skills to program, not to master some specific syntax / flavour of programming.

\vspace{-0.5cm}
\section{The Program}
\label{sec:program}

With some small modifications depending on the year, the program of LSSDS consists roughly of 34\% of the time for lectures, 23\% of the time for Hands-on / Lab sessions, and 43\% of the time allocated to ``project" work.

The lectures cover a variety of topics: Principles of data science and statistics, software carpentry, machine learning (``classic” and deep learning architectures), databases, cloud and parallel data access and computing, image processing (of astronomical or bio-sciences origin), time series data-analysis, and, for the past years, large language models and natural language processing (NLP). Most lecture material comprises of slides and python notebooks and are provided to the students either during the lessons or in advance (see below). The teaching style varies among lectures, but is heavily influenced by active learning methods.

The hands-on sessions (sometimes inserted among the lectures) consist in most cases of python notebooks with missing cells of code (or cells with errors) to solve problems derived from the material explained in the lectures. The students are encouraged to discuss with each other to solve the problems, as well as with faculty and the Teaching Assistants (TAs, a component that we only incorporated during the virtual editions, see below).

On day one of the school, research projects are presented by the faculty to the students (``elevator pitch style"). The goal is that only faculty that can be available through the whole school propose research projects, however some compromises have had to be made on a few occasions; in those cases we encourage several faculty members to team up to coordinate adequate supervision through the whole school. Right before these project introductions, groups of four students are semi-randomly assigned (to balance gender, field of studies and nationality). By the end of the day, the groups have time to meet, start to know each other, and decide on a ranking of projects according to their interests. 

On the second day, the projects are assigned to the groups, and the project work time increases as the school progresses (with the second half fully devoted to it). The selection of projects is wide enough that normally we can accommodate fully the preferences of the groups and assign them to a project in their top three options. Once the groups start working on the projects we reserve the last $\sim$30 mins of each day for a project representative to very briefly present the progress. These instances are very informal, the group's representative has to change every day, and they are used to provide some training on effective communication, and, in general, soft skills. Circling back to diversity, we must note that for many of the Chilean students these instances are among the first opportunities they have to present in English. And, although we have witnessed instances of struggle, the feedback that we get constantly is that these short reporting instances are extremely beneficial for their future. 

Although the second half of the school does not include lessons in the schedule, the opportunity is given to the students to propose optional lectures. For instance, before NLP was one of the core topics, it was a topic often requested by students. 

The end of the school is devoted to group presentations on the results of their projects followed by discussions where students from other groups are encouraged to ask questions.

In part because of the large oversubscription of the school and also for the general commitment by the faculty and organizers to Open Science, an effort has been pursued every year to make the material fully available to the general community: until 2023 this was done via the school webpage (under the ``Final Program" tab), and in the last edition the TAs took the initiative to create a github repository.\footnote{\url{https://github.com/lssds2024/lssds2024}}

\vspace{-0.3cm}
\subsection{``Regular", in-person, vs virtual editions:}

A very important aspect of the ``regular", in-person experience, other than the constant direct contact among students and faculty, is the visit to the AURA telescopes, including a stargazing event, that ``breaks" the school after five very intense days of learning.

On the other hand, while this exciting and bonding moment could not be achieved during the virtual editions, the LOC made huge efforts to generate a sense of belonging in the latter. In particular, they prepared a very complete physical welcome packages that where shipped to all students and included not only a reference book but also ``comfort goodies" like coffee, tea and chocolate to be virtually shared during the breaks.

In addition, to account for the many different time zones of the participants, the lecture time was divided into asynchronous (with pre-recorded short videos) and synchronous time, and the platforms used included Zoom, Slack and Gather Town.

Last, but not least, during the virtual editions, former LSSDS students (alumni) were recruited to help running the school as TAs. This experience was extremely positive for all parties (new students, the faculty and the TAs themselves) and we kept this feature in the last two in-person editions.

\vspace{-0.5cm}
\section{Lessons Learned \& Legacy}
\label{sec:lessons}

Over the years, LSSDS has incorporated some modifications. The inclusion of TAs, as previously mentioned, was one of the very successful ones, while others did not provide satisfactory results. Among the latter, we can highlight the inclusion of a \textit{Pre-school Coding Bootcamp} and the presence of highly specialized statistical / thematic lectures among the core topics. The former was offered to level the coding skills of the participants, however, the general feeling was that in fact exacerbated the differences in skills rather than level. The latter ``clashed" with the generic approach of the school and was morphed into ``on-demand" lectures that the students could attend during project time.

After ten editions of LSSDS, a note has to be made on the legacy value. Besides the constant very positive feedback left by the students in our yearly survey, we would like to highlight the fact that alumni were thrilled to ``come back" as TAs, that some alumni even returned, years after, as faculty, and a number of observing proposals and scientific publications were enabled by the school \citep{Cabrera17, Forster22}.

\vspace{-0.5cm}
\section{A Different Approach to Similar Needs}
\label{sec:SVO}
Another long-lasting hands-on teaching initiative (operating since 2009) is the series of schools led by the Spanish Virtual Observatory (SVO\footnote{\url{https://svo.cab.inta-csic.es/docs/index.php?pagename=Meetings}}). These are shorter, much more ``VO astronomy-tool" focused schools that explore diversity by being open to both professional and amateur astronomers. The learning of the tools is acquired by working on pre-defined science cases and, since 2020, the organizers also made the transition to combine in-person and virtual editions. 


\vspace{-0.5cm}
\begin{small}
\section*{Acknowledgements}
Thanks to funding from NSF (Award Number (FAIN): 1637359)  applications from students in US institutions are eligible for full scholarships that cover all school expenses.

\begin{discussion}
\discuss{Q1}{I'm wondering how do you include soft-skill teaching with all the technical training already in the school?}
\vspace{-0.1cm}
\discuss{A1}{The short answer is that we do not do formally, but we do it via the daily report and ``enforcing" the rotation of the role of presenter to account for cultural differences between US and Chilean based students.}
\vspace{-0.1cm}
\discuss{Q2}{Has the knowledge expressed in the lessons learned part been published / preserved further than by the SOC / LOC?}
\vspace{-0.1cm}
\discuss{A2}{Unfortunately no, but the proceeding of this meeting can be a good place for that.}
\end{discussion}
\end{small}

\end{document}